\documentclass[aps,prb,reprint,superscriptaddress]{revtex4-1}

\usepackage{graphicx}
\usepackage{subfigure}
\usepackage[ugly]{nicefrac}
\usepackage{amsmath}
\usepackage{amsthm,color}
\usepackage{comment}
\usepackage[toc,page]{appendix}

\begin{document}


\title{Effects of electron-hole asymmetry near the Dirac point in graphene.}


\author{Salvatore Croce}
\affiliation{Department of Mathematical Sciences, Durham University,
South Road, Durham DH1 3LE, United Kingdom}


\date{\today}

\begin{abstract}
In the recent years   many researches were performed about graphene,  as we can see from \cite{Novoselov-2005,Geim-2007,PhysRevLett.97.266801,Dean-2011}. Graphene is always considered a half metal or a zero gap semiconductor. In the last year new experiments were  done about graphene on boron nitride\cite{Dean-2010}, and in particular in a work they obtained an insulating behaviour and a power law for conductivity at the charge neutrality point (CNP) or Dirac point (DP)\cite{PhysRevLett.110.216601}. In this work we will explore the electron-hole asymmetry and the disorder in graphene. In particular we  analyze the asymmetry in the ordered phase, considering  thermal doping at half filling and with a non zero chemical potential. After that we  use the self consistent Born approximation for calculating the self energy correction from impurity scattering. \\
Furthermore we shall analyze the electron-phonon interaction, using the deformation potential model for obtaining the temperature dependence for the compressibility. Finally we will speculate about temperature dependence of conductivity.
\end{abstract}

\pacs{}
\keywords{graphene, disorder, conductivity}

\maketitle

\section{Introduction}
Graphene is a mono layer material  with a honeycomb lattice made by carbon atoms and it is a relative of graphite. It can be considered the first two dimensional material, since its discovery in 2004\cite{Novoselov22102004}. In fact Geim and Novoselov discovered how to stabilize a two dimensional material by means of a substrate, against the general opinion motivated by the Mermin-Wagner theorem \cite{PhysRev.176.250}. In the last years many scientific works were dedicated to Graphene like \cite{RevModPhys.81.109}  where we find many of the electronic properties and \cite{Das-2008,10.1038/nature04235} about experimental properties, because graphene is a zero gap semiconductor or a half metal with a very peculiar band structure near the Fermi level\cite{PhysRev.71.622} with a linear dispersion instead of a parabolic one like many other semiconductors. This properties bring with them many peculiar and fascinating characteristics that can be used in applications \cite{Novoselov-2005}. Up to 2013 All the studies were performed on graphene on $Si/SiO_2$ but it was pointed that this kind of substrate can modify the properties of graphene with a modification of the Fermi velocity and the band structure\cite{10.1038/nphys2049}. Just in the last two years new experiments were performed on a substrate of boron nitride, a two dimensional material with atoms of boron and nitride arranged in a honeycomb structure but with a semiconducting behaviour Furthermore all previous experiments where performed with a single gate set up, checking just the doping of the system. In a recent experiment\cite{PhysRevLett.110.216601} They performed measures on graphene  deposed on boron nitride with a double gate experiments which allows them variations of the number of particles and also the chemical potential. Furthermore new technology advances brought to the market new organic gates with an increase efficiency.  They discovered an insulating behaviour near the CNP and a very strong temperature dependence of the conductivity as a power law. In this work we aim to derive some important properties of graphene near the Dirac point and  the effects of the electron-hole asymmetry. In particular in the second section we shall analyze the e-h asymmetry on the density of states and electronic properties. In the third section we will see the thermal doping in the presence of the asymmetry in the non interacting system. In the fourth section we will see the role of the static and dynamical disorder by means of the Self Consistent Born Approximation (SCBA) for impurity scattering and the electron-phonon interaction and the derivation of the self energy and the interacting DOS with the spectral properties. Moreover we will analyze the effects of the disorder in the Dirac point and finally in the last section we shall see the compressibility of the system that may be linked to a quantitative explanation of the anomal conductivity.  
\section{Low energy electronic model and e-h asymmetry}
In this section we are going to introduce our low energy model for describing the electron-hole asymmetry. As we have seen so far the graphene's band structure near the Fermi level presents itself like two cones touching in one point and the density of states is a linear function of the energy with a different sign for electrons and holes as we can see from \cite{PhysRev.71.622}. In particular we will calculate a correction to the bare density of state (DOS) in the Dirac point (DP), obtaining a non linear behavior. In fact if we consider a Taylor expansion of the band structure up to the second order, as obtained by\cite{RevModPhys.81.109}, we have:
\begin{equation}
\label{svilrid}
E_{\pm}(q,\theta)=3t^{'}\pm v_{f}q- A(\theta)q^2
\end{equation} 
where $A(\theta)=\frac{9t^{'}a^2}{4}\pm  \frac{3ta^2}{8}\cos[3\theta]$. If we use  the Jacobian of the transformation and the lower and upper values of $q$ and $\theta$ the DOS is \label{polardos}
$\rho(E)=\frac{L^2}{(2\pi)^2}\int_{0}^{2\pi}d\theta \int_{0}^{\infty}\delta(E-E_{\pm}(q,\theta))q d q$, where  $L^2=|\boldsymbol{a_1}\times \boldsymbol{a_2}|$ is the volume of the unitary cell. The calculation can be simplified by means of a shift of the energy variable as $\epsilon=E-3t^{'}=0$ and dividing the integral between electrons and holes, where we use a plus sign at the right of the DP and a minus one at the left. We notice that in the limit of $t^{'} \to 0$ we obtain again the  expression of a symmetric DOS. Furthermore we shall write the DOS in a more compact form as:
\begin{eqnarray}
\label{compactdos}
\rho_{\pm}(\epsilon)=(\alpha |\epsilon|+\mbox{sgn}(\epsilon)\beta \epsilon^2)
\end{eqnarray}
where $\alpha=\frac{3\sqrt{3}a^2}{ 4\pi}\frac{1}{v_{f}^2}$  and $\beta=\frac{27\sqrt{3}a^3}{8 \pi}\left(\frac{t^{'}}{t}\right)\frac{1}{v_{f}^3}$. Using a value of $t^{'}=0.2t$ as reported in literature\cite{review electronic properties}, we obtain the numerical values (in hopping units)  $\alpha \simeq 0.368$ and $\beta \simeq 0.221$.
\section{Thermal doping in non-interacting system}
In this section we are going to derive the thermal doping in the non interacting graphene. As we have seen so far the new experiments concerning the double gate doping in nano-structures are suitable to control not only the number of particles but also the chemical potential, so it might be important to have an analytical expression for that. As previously pointed out by Guinea et al. \cite{PhysRevB.88.165427} we have thermal doping just with electron-hole asymmetry. In particular we may have three different cases, distinguished by the values of the chemical potential:
\begin{enumerate}
\item The DP with $\mu=0$, analyzing the number of particles as a function of temperature;
\item The case with a non zero chemical potential $\mu \neq 0$ which can be split in two sub-cases of low and high temperature:
\begin{itemize}
\item the low temperature limit $|\mu|<T$ that is equal to the DP thermal doping plus some corrections derived by an expansion of the Fermi-Dirac function; 
\item the high temperature limit $|\mu|>T$ where it is possible using the Sommerfeld expansion with properly used cut-off;
\end{itemize} 
\end{enumerate}
\subsection{Thermal doping at the Dirac point}
If we fix the chemical potential in a double gate experiment we are able to use the symmetry of the Fermi-Dirac function for evaluating the thermal doping. In fact dividing the DOS in a symmetric part (first order in $\epsilon$) and a non symmetric part (second order in $\epsilon$) adding plus and minus $\frac{1}{2}$ to the expression for the number of particles we have:
\begin{equation}
 \label{nummuzero}
n(\mu,T)=\frac{1}{2}\int_{-\infty}^{+\infty}(\rho_s(\epsilon)+\rho_{as}(\epsilon))(1-\tanh{\left(\frac{\epsilon-\mu}{2T}\right)})d \epsilon
\end{equation}
where  $\rho_s(\epsilon)=\alpha|\epsilon|$ e $\rho_{as}(\epsilon)=\mbox{sgn}(\epsilon)\beta \epsilon^2$. We notice that the first integral is zero, because the function is odd in an even interval, so we shall solve just the second one.
Using a simple substitution $\frac{\epsilon}{2T}=x$, we obtain the final expression with a power law function for the temperature in the DP:
\begin{equation}
n(\mu=0,T)-n(\mu=0,T=0)=8T^3\beta I_2
\end{equation}
with $I_2=\int_{0}^{+\infty} x^2(1-\tanh{x})dx=\frac{3}{8}\zeta(3)$, with the numerical value:
\begin{equation}
\label{muzero}
n(T)-n_0=3.61\beta T^3
\end{equation}
where the number of particles for a half filling not doped system is $n_0=2$ with a maximum equal to $4$. We notice that for $\beta \to 0$, i.e. $t^{'} \to 0$ we do not have a temperature dependence of the number of particles, because it is a direct consequence of the electron-hole asymmetry of the density of states.
\subsection{Non-zero chemical potential}
The low temperature regime is characterized by $\mu \neq 0$ and $\mu \gg T$ and with these conditions we can use the Sommerfeld expansion. Following the technique reported in literature \cite{ashcroft, grosso2000solid} we can write the number of particles with both contributions from electrons and holes as:
\begin{equation}
n(\mu,T)=\int_{-\infty}^{0}\rho_{-}(\epsilon)f(\epsilon)d \epsilon+\int_{0}^{+\infty}\rho_{+}(\epsilon)f(\epsilon)d \epsilon
\end{equation}
where $\rho_{-}$ and $\rho_{+}$ refer to the density of states at left and right of the DP, respectively. In the limit of low temperature and expanding the integrals in a series:
\begin{equation}
\label{som}
n(\mu,T)-n_0=sgn(\mu)\left[\frac{\alpha}{2}\mu^2+\frac{\beta}{3}\mu^3+\frac{\pi^2}{6}T^2(\alpha+2\beta \mu)\right]
\end{equation}
where we see that is a small corrections to the thermal doping induced by $\beta$ and for $T \to 0$ we obtain the integral of the density of states of the system.
The final expression in the high temperature limit is:
\begin{equation}
\label{mumint}
\begin{split}
&n(\mu,T)-n_0=\\
&3.61\beta T^3+\mu \left(2 \alpha TI_1+2 \beta T^2I_2 \right)-\mu^2\left(  \alpha I_0+2\beta TI_1\right)
\end{split}
\end{equation}
where we notice a symmetry behaviour for $\mu \to 0$, with an expression given again by eq.(\ref{muzero}). We see the eq.(\ref{mumint}) as a function of temperature and the chemical potential in fig.(\ref{numbsom}) with a comparison with the exact numerical result. The arrow set the limit of validity of the analytical solution. 
\begin{figure}[htbp]
\begin{center}
\subfigure{
\includegraphics[width=0.8\columnwidth]{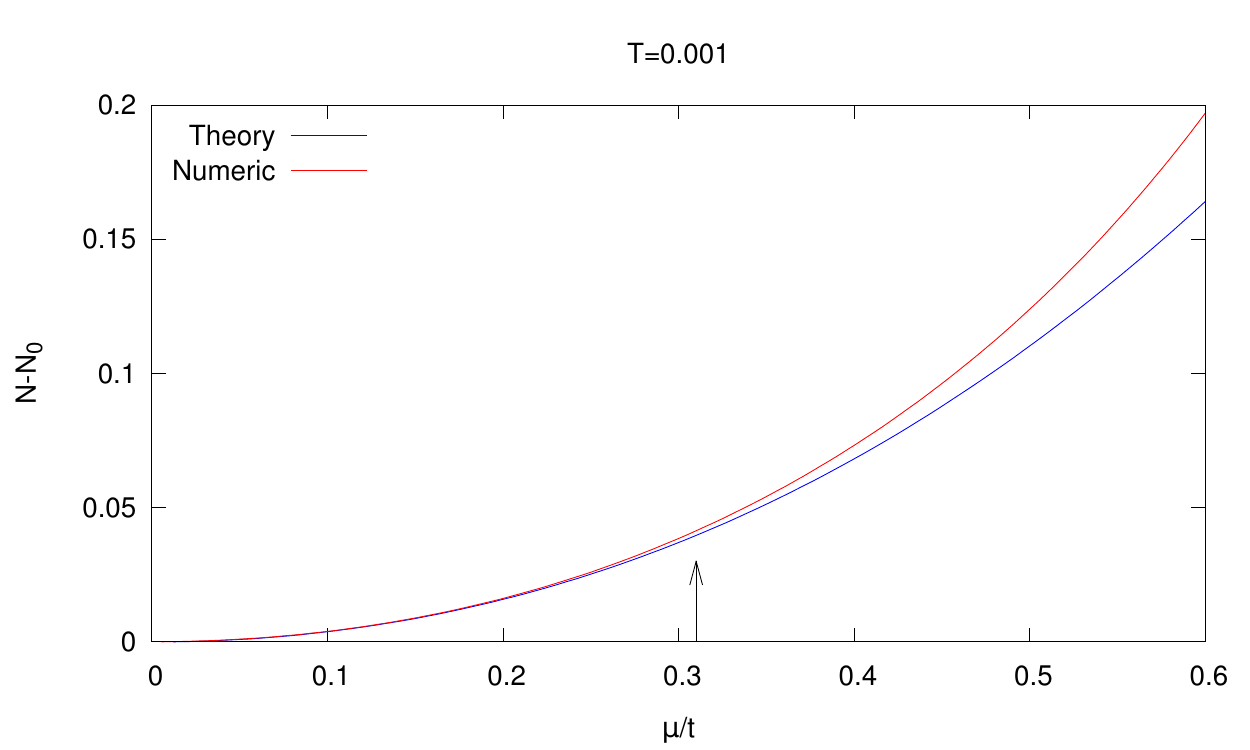}}
\subfigure{
\includegraphics[width=0.8\columnwidth]{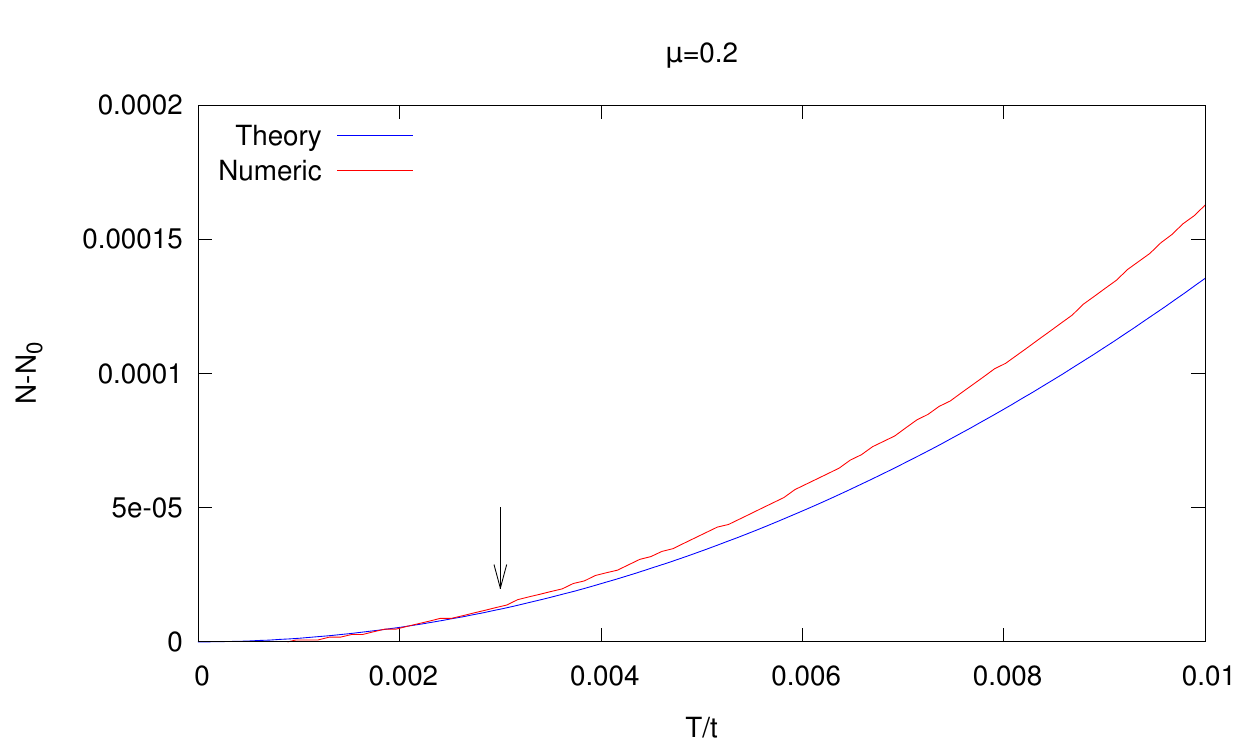}}
\caption{Particle number as a function of the chemical potential (fixed temperature) and Temperature (Fixed chemical potential). In both pictures $k_b=1$, the energy scale is in hopping units  $t^{'}=0.2t$ and everything is shifted by $n_0=2$}
\label{numbsom}
\end{center}
\end{figure}
%

\section{Role of static and dynamical disorder}
\subsection{SCBA + el-ph interaction}
In this section we are going to see the effects of the interactions on the system. In fact we have seen all the previous results with a zero temperature and without interactions. In particular following previous works we can consider the effects of impurities by means of the Self consistent Born approximation (SCBA) \cite{JPSJ.67.2857,JPSJ.67.2421} and also the  combination of the SCBA and the electron-phonon interaction\cite{PhysRevB.68.224511}. In particular if we consider a potential for short-range scatterers like in Ando\cite{JPSJ.67.2857},with the Dyson equation and the SCBA the Green's function will be a diagonal matrix, indipendent from $k$ and the self energy can be written as a product of the averaged potential with the averaged Green function with a Gaussian disorder, so:
\begin{equation}
\label{auto}
\begin{split}
&\Sigma (\epsilon)=\\
&n_i \sigma^2 \left( \int_{-\Lambda_{-}}^{0}d \epsilon^{'} \frac{\rho_{-}(\epsilon^{'})}{\epsilon-\epsilon^{'}-\Sigma(\epsilon)}+\int_{0}^{\Lambda_{+}}d \epsilon^{'} \frac{\rho_{+}(\epsilon^{'})}{\epsilon-\epsilon^{'}-\Sigma(\epsilon)}  \right)
\end{split}
\end{equation}
where  $\Lambda_{+}$ e $\Lambda_{-}$ are integration cut-off, different because the DOS is asymmetric, $\sigma^2$ is the variance of the Gaussian disorder and $n_i$ the density of scatterers.
 Considering also the electron-phonon interaction we can write the self energy of the electron as:
\begin{equation}
\label{selfelectron0}
\begin{split}
&\Sigma(i\omega_n+i\nu_m,\mathbf{k})=\\
&=\frac{1}{\beta}\sum_{m,\mathbf{q}}|g_{\mathbf{q}}|^2D(\mathbf{q},i\nu_m)G(\mathbf{k+q},i\omega_n+i\nu_m)
\end{split}
\end{equation}
where the bosonic propagator is:
\begin{equation}
\label{boson}
D(\mathbf{q},i\nu_m)=-\frac{2\omega_{\mathbf{q}}}{\omega_{\mathbf{q}}^2+\nu_m^2}
\end{equation}
and the Fermionic propagator:
\begin{equation}
\label{propgreen}
G(\mathbf{k+q},i\omega_n)=\frac{1}{i\omega_n+\mu-\epsilon_{\mathbf{k}}-\Sigma(i\omega_n,\mathbf{k})}
\end{equation}
and $\omega_n$ and $\nu_m$ are respectively the Fermionic and Bosonic frequencies as reported in Mahan\cite{Mahan}.\\
Considering the approximation of classical phonons $\hbar \omega_{\mathbf{q}}\ll k_b T$ we can still write the self energy as independent   from $k$, like in the SCBA from impurity scattering, with a propagator given by:
\begin{equation}
D(\mathbf{q},i\nu_m) \simeq -\delta_{m,0}\frac{2}{\omega_q}
\end{equation}
and so the Self energy, a totally isotropic function in the 1Bz, is 
\begin{equation}
\Sigma(i\omega_n,\mathbf{k})=T\sum_{\mathbf{q}}\int_{0}^{\infty}d\omega 2\frac{\alpha^2F(\mathbf{q},\omega)}{\omega}G(\mathbf{k+q},i\omega_n)
\end{equation}
where $\alpha^2F(\mathbf{q},\omega)=I |\mathbf{q}|\delta(\omega-\omega_{\mathbf{q}})$ is the Eliashberg function and $I$ is linked to the deformation potential as reported by Mahan \cite{Mahan}. In this way we obtain:
\begin{equation}
\label{selfloc}
\Sigma(i\omega_n,\mathbf{k})\simeq \sigma^2 G_{\mbox{loc}}(i\omega_n)
\end{equation}
where the variance of the disorder is the same of the variance obtained for the impurity scattering. If we write the variance by means of the microscopic parameters of the e-ph interaction, with the deformation potential we have:
\begin{equation}
\sigma^2=\frac{2T I}{\hbar v_s}=T\frac{D^2}{V\rho_m v_s^2}
\end{equation}
with $D$ deformation constant, $v_s$ the sound velocity, $\rho_m$ the density and $V$ the volume of the unitary cell. Using the values reported in literature \cite{,PhysRevB.79.035419} we obtain:
  \begin{equation}
\label{tempdis}
\frac{\sigma^2}{t^2}=1.3(\frac{k_bT}{t})
\end{equation}
which is the variance of the disorder for $t^{'}=0.2t$ and it is a linear function of the temperature.
\subsection{Energy dependent properties}
We have seen that we can obtain an expression of the self energy of the system by means of eq.(\ref{auto}). and finally we arrive to an expression for the self energy which includes the effect of the e-h asymmetry:
\begin{equation}
\label{auto2}
\begin{split}
&\frac{\Sigma (\epsilon)}{n_i \sigma ^2}=\alpha \left( \Lambda_{-}-\Lambda_{+}+ z\ln{\frac{z^2}{(z+\Lambda_{-})(z-\Lambda_{+})}}  \right)+\\
&+\beta \left( zG(z)-\frac{\Lambda_{+}^2+\Lambda_{-}^2}{2} \right)
\end{split}
\end{equation}
where we notice that using the substitution $z=\epsilon-\Sigma(\epsilon)$ we obtain the self consistency and with the limit $\beta \to 0$, also $\Lambda_{+}=\Lambda_{-}$ and we obtain a similar expression like in Ando\cite{JPSJ.67.2421}.\\
If we neglect the self consistency in eq.(\ref{auto2}) we can obtain an analytical expression for the self energy of the system and analyzing the spectral properties. In particular using eq.(\ref{noncons2}) we can derive the expressions of the real ($\Delta$) and imaginary ($\Gamma$) parts of the self energy:
\begin{equation}
\label{analiticself}
\begin{split}
&\Gamma(\omega)=-n_i \sigma^2\pi\mbox{sgn}(\omega)(\alpha \omega+\beta \omega^2)\\
&\Delta(\omega)=n_i \sigma^2[ (\alpha+\beta \omega) ( \Lambda_{-}-\Lambda_{+}+ \\
&+\omega\ln{\left(\frac{-\omega^2}{(\omega+\Lambda_{-})(\omega-\Lambda_{+})}\right)})-\beta \frac{\Lambda_{+}^2+\Lambda_{-}^2}{2}]
\end{split}
\end{equation}
that are expressions valid inside the energy between the cut-off. We notice the minus sign of the imaginary part of the self energy, needed because it is directly related to the density of states and again we obtain the symmetric version imposing the symmetry as $\beta \to 0$. If we change the sign of the imaginary part we can compute the spectral DOS as:
\begin{equation}
\begin{split}
&DOS(\omega)=-\frac{1}{\pi} [\int_{\Lambda_{-}}^{0}d\epsilon \rho_{-}(\epsilon)\mbox{Im}\frac{1}{\omega-\epsilon-\Sigma(\omega)}+\\
&+\int_{0}^{\Lambda_{+}}d\epsilon \rho_{+}(\epsilon)\mbox{Im}\frac{1}{\omega-\epsilon-\Sigma(\omega)} ]
\end{split}
\end{equation}
and so we obtain an expression for the interacting density of states:
\begin{equation}
\label{dosint}
\begin{split}
&DOS(\omega)=\\
&-\frac{1}{\pi} \mbox{Im}[ (\alpha (\Lambda_{-}-\Lambda_{+}+z \log{\frac{z^2}{(z+\Lambda_{-})(z-\Lambda_{+})}})+\\
&+\beta z(\Lambda_{-}-\Lambda_{+}+z \log{\frac{z^2}{(z+\Lambda_{-})(z-\Lambda_{+})}})-\\
&-\beta \frac{\Lambda_{+}^2+\Lambda_{-}^2}{2}]
\end{split}
\end{equation}
where $z=\omega-\Delta(\omega)+i\Gamma(\omega)$ with $\Delta(\omega)=\mbox{Re}\Sigma(\omega)$ and $\Gamma(\omega)=\mbox{Im}\Sigma(\omega)$.  Considering the different Riemann surfaces we have the final expression:
\begin{equation}
\begin{split}
&DOS(\omega)=\\
&-\frac{1}{\pi}\{ (\alpha\Gamma(\omega)+2\beta\Gamma(\omega)(\omega-\Delta(\omega)))\log{\frac{\rho_1}{\sqrt{\rho_2\rho_3}}}+\\
&+\mbox{sgn}(\omega)(2\theta_1-\theta_2-\theta_3)\\
&[\alpha(\omega-\Delta) +\beta((\omega-\Delta(\omega))^2-\Gamma(\omega)^2)]+\\
&+\Gamma(\omega)\beta(\Lambda_{-}-\Lambda_{+})\}
\end{split}
\end{equation}
where the $\rho_i$ and $\theta_i$ are functions of $\omega$ as reported in the appendix.
In the next section we are going to analyze what are the properties spectral properties in the DP.
\subsection{Dirac point properties}
\begin{figure}[h]
\begin{center}
\includegraphics[width=1.0\columnwidth]{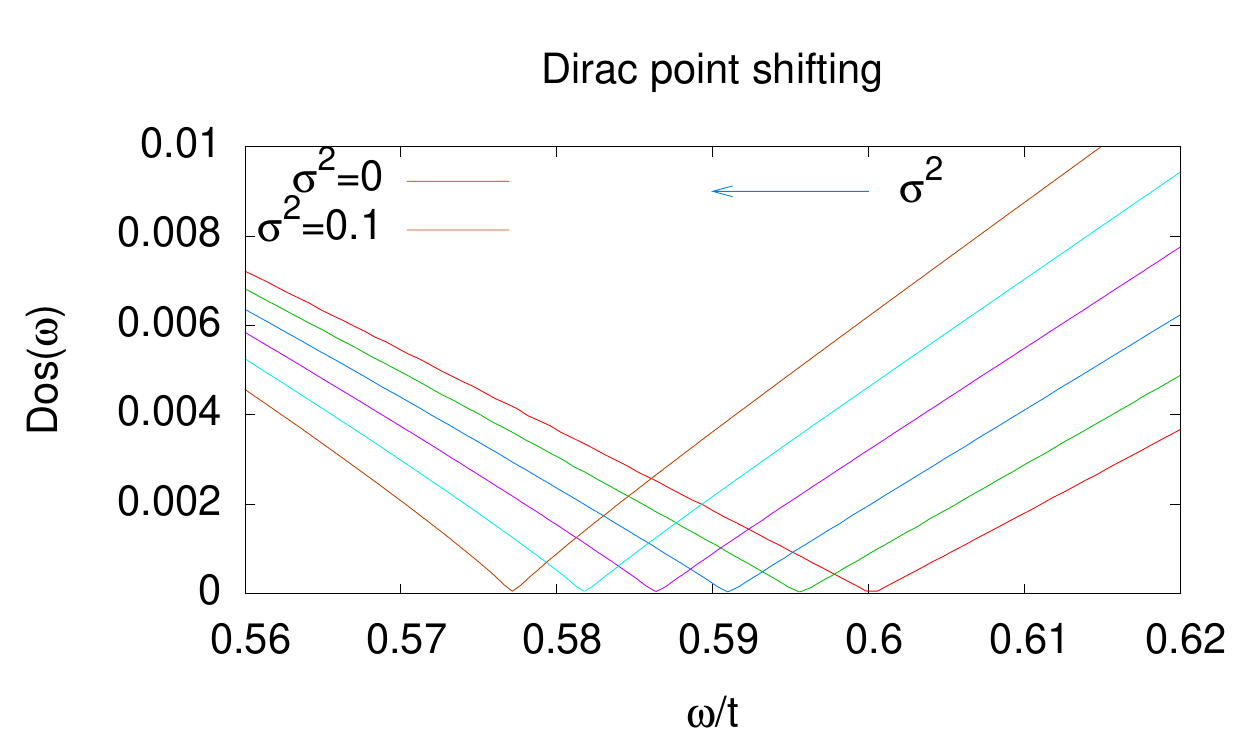}
\caption{Continuous shifting with increasing disorder. It is a numerical calculation with $\sigma^2$ in the range from $0$ to $0.1$, with an increment of $\sigma^2=0.02$. The picture is in hopping units with $t^{'}=0.2t$}
\label{confrontoselfdos2}
\end{center}
\end{figure}

We shall be able to analyze the disorder corrections to the Dirac point starting from eq.(\ref{auto2}) at fixed energy,i.e. the DP as $\Delta=\mbox{Re} \Sigma=\omega$ and so $\Gamma=\mbox{Im} \Sigma$. In this condition the real part is a correction to the electronic energies, a band renormalization, and the imaginary part is a correction to the density of states of the system. In this condition the self energy will be $\Sigma=-i\Gamma$ and so  $z=i\Gamma$. Remembering the expressions of the functions   in eq.(\ref{fun_self}) we obtain:
\begin{equation}
\label{imag}
\begin{split}
&G(i\Gamma)=(\Lambda_{-}-\Lambda_{+})+\\
&+i\Gamma \ln{\left[\frac{-\Gamma^2}{(i\Gamma+\Lambda_{-})(i\Gamma-\Lambda_{+})}\right]}\\
&F(i\Gamma)=i\Gamma G(i\Gamma)-\frac{\Lambda_{-}^2+\Lambda_{+}^2}{2}
\end{split}
\end{equation}
Now at fixed energy we can neglect the contribution of $\Gamma$ inside the logarithm, because $\Lambda_{+},\Lambda_{-} \gg \Gamma$, so:
\begin{equation}
G(i\Gamma) \simeq (\Lambda_{-}-\Lambda_{+})+i\Gamma \ln{\left[\frac{\Gamma^2}{\Lambda_{-}\Lambda_{+}}\right]}
\end{equation}
and the final equations will be:
\begin{equation}
\label{self_dirac}
\begin{split}
&\frac{\Sigma}{n_i\sigma^2}=[(\alpha +i\beta \Gamma) \left(\Lambda_{-}-\Lambda_{+}+i\Gamma \ln{\left[\frac{\Gamma^2}{\Lambda_{-}\Lambda_{+}}\right]}\right)-\\
&-\beta \frac{\Lambda_{-}^2+\Lambda_{+}^2}{2} ] 
\end{split}
\end{equation}
which is an explicit solution of eq.(\ref{auto2}) at fixed energy $\omega=\Delta=\mbox{Re} \Sigma$.  \\
Considering separately the contribution of the imaginary part, from eq.(\ref{self_dirac}) we obtain:
\begin{equation}
\frac{-\Gamma}{n_i\sigma^2}=  \Gamma\left( \beta(\Lambda_{-}-\Lambda_{+})+ \alpha \ln{\left[\frac{\Gamma^2}{\Lambda_{-}\Lambda_{+}}\right]}\right)
\end{equation}
and in the same limit of $\Lambda_{-},\Lambda_{+} \gg 1$ we have an exponential small correction to the density of states which is different from zero at the DP:
\begin{equation}
\label{gapdos}
\Gamma=\sqrt{\Lambda_{+}\Lambda_{-}}e^{-\frac{1}{2\alpha}(\frac{1}{\sigma^2}+\beta(\Lambda_{-}-\Lambda_{+}))}
\end{equation}
we notice that in the limit of $\beta \to 0$ the eq.(\ref{gapdos}) is the same of Ando\cite{JPSJ.71.1318}, because $\Lambda_{-}=\Lambda_{+}$. With a perfect analogy we can define a renormalized disorder:
\begin{equation}
\sigma_{*}^2=\frac{\sigma^2}{1+\beta\sigma^2(\Lambda_{-}-\Lambda_{+})}
\end{equation}
with a variance of the effective disorder that can be increased if $\beta$ has a negative sign and it can be decreased if $\beta$ has a positive sign, because $\Lambda_{-}>\Lambda_{+}$ as we will see later. This is a totally new results and it is important its connection to both the static and dynamic disorder. In fact if we consider a system with impurities, combining the effect with the e-h asymmetry we can increase the effect of the renormalization on the DOS and in a system with dynamic disorder and a scattering from phonons, we can increase the effect of the temperature to the DOS using the eq.(\ref{tempdis}).\\
Furthermore we shall consider the effect of the real part of the self energy, from eq.(\ref{self_dirac}) we have:
\begin{equation}
\frac{\Delta}{n_i\sigma^2}=\left[ \alpha(\Lambda_{-}-\Lambda_{+}) -\beta \Gamma^2 \ln{\left[\frac{\Gamma^2}{\Lambda_{-}\Lambda_{+}}\right]}- \beta \frac{\Lambda_{+}^2+\Lambda_{-}^2}{2}\right]
\end{equation}
  and using the previous result $ \ln{\left[\frac{\Gamma^2}{\Lambda_{-}\Lambda_{+}}\right]}=-\frac{1}{\sigma_{*}^2 \alpha}$ and being multiplied by $\Gamma^2$ thanks to eq.(\ref{gapdos}) it will be exponentially small and it can be neglected. Defining $\Lambda_{-}-\Lambda_{+}=\frac{\beta}{\alpha} \delta$ We obtain:
\begin{equation}
\label{om0}
\Delta=n_i \sigma^2\beta(\delta-\frac{\Lambda_{+}^2+\Lambda_{-}^2}{2})
\end{equation}
which is an amazing result, because it introduces a linear dependence of the  Dirac point  with the variance of the disorder $\sigma^2$ and it moves to negative energies increasing $\sigma^2$. We deduce that the disorder can shift the Dirac point. Taking into account the  low disordered nature of Graphene, the effect it will have a physical meaning just for a value of the variance up to $\sigma^2 \sim 0.1 $. In particular also in this case if we consider just the static disorder we can shift and fix the position of the DP, fixing the variance of the impurities in the system and considering again eq.(\ref{tempdis}) we can also move the DP linearly with the temperature in a continuous way, managing the position of the gap of the system with the temperature. In particular if we substitute the numerical values of the constants in eq.(\ref{om0}) we obtain:
\begin{equation}
\frac{\Delta}{t}=1.11\beta(\frac{k_bT}{t})
\end{equation}
And in the particular case of $\ t^{'}=0.2t$ we obtain:
\begin{equation}
\frac{\Delta}{t}=0.25(\frac{k_bT}{t})
\end{equation}
We can see those effects of eq.(\ref{om0}) and eq.(\ref{gapdos}) in fig.(\ref{confrontoselfdos2}), where we observe both the shift and the non zero value of the DOS.

\begin{figure}[htbp]
\begin{center}
\includegraphics[width=0.8\columnwidth]{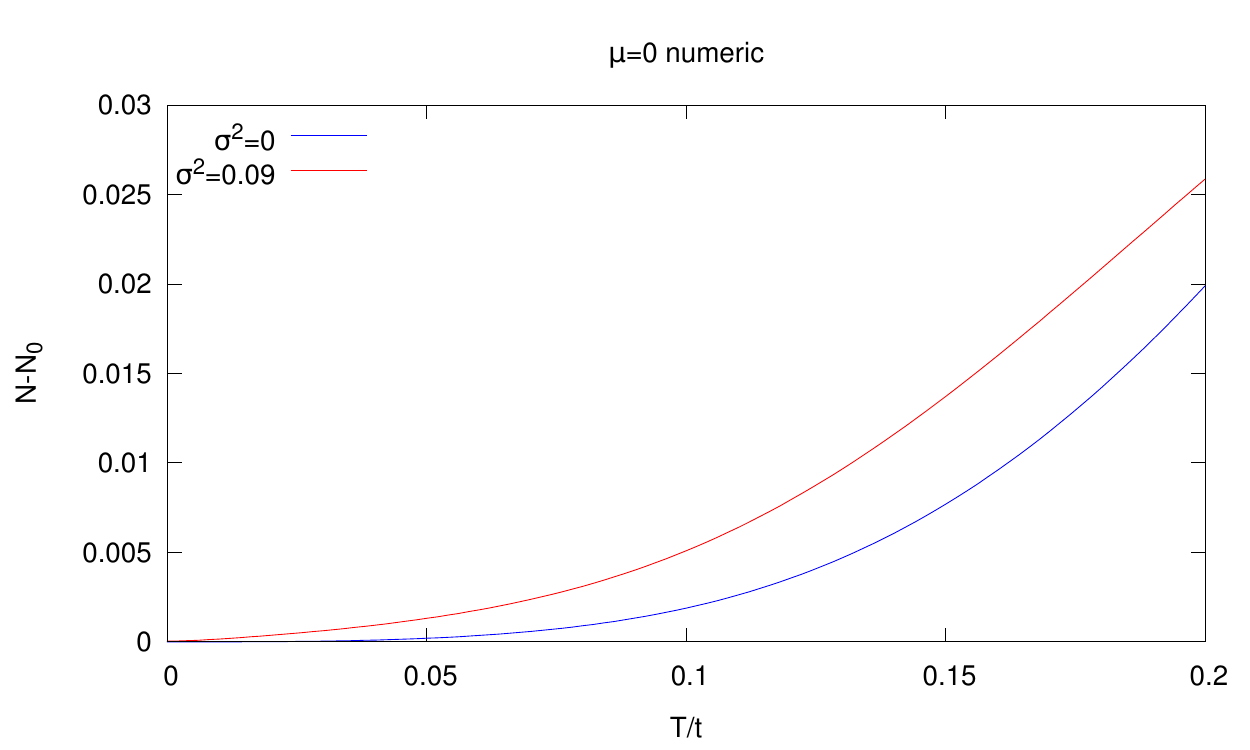}
\includegraphics[width=0.8\columnwidth]{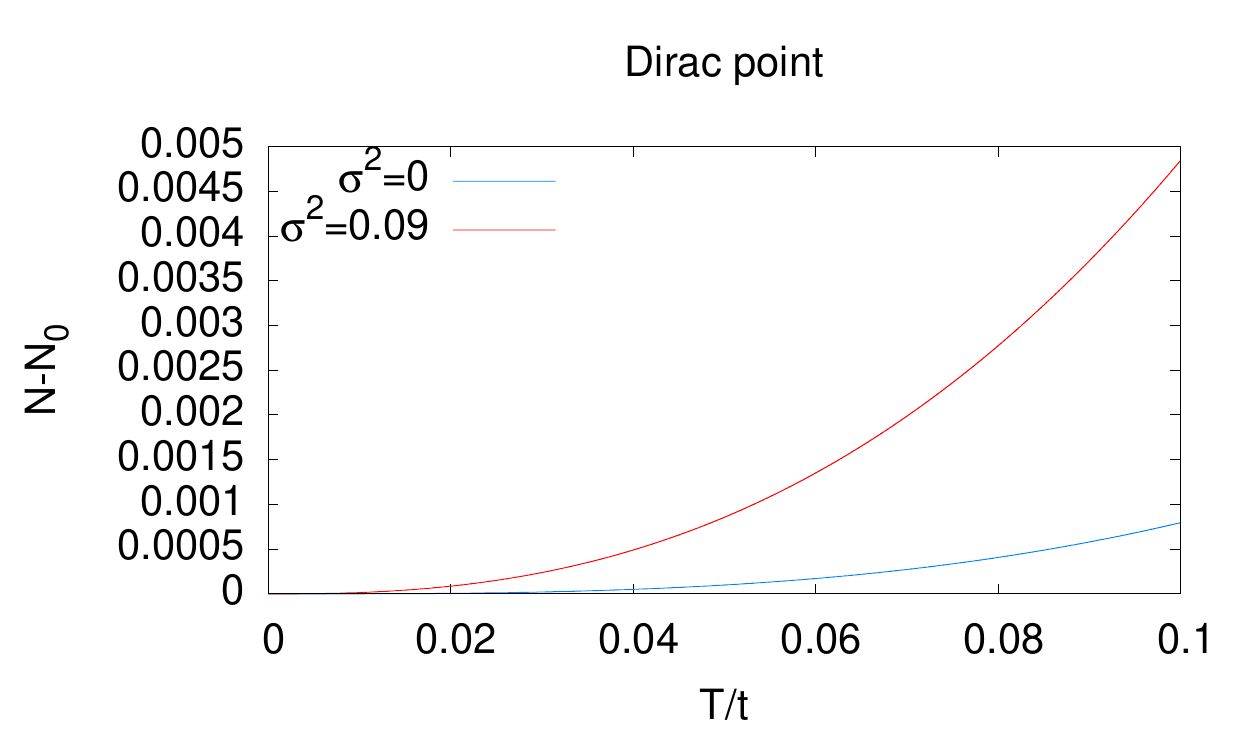}
\caption{Termal Doping in the Dirac point with and without disorder calculated numerically (top) and comparison between the analytic solutions without disorder and a fit with fixed disorder (bottom). The picture is in hopping units with $t^{'}=0.2t$ and $n_0=2$}
\label{particleself0}
\end{center}
\end{figure}

\begin{figure}[htbp]
\begin{center}
\includegraphics[width=0.8\columnwidth]{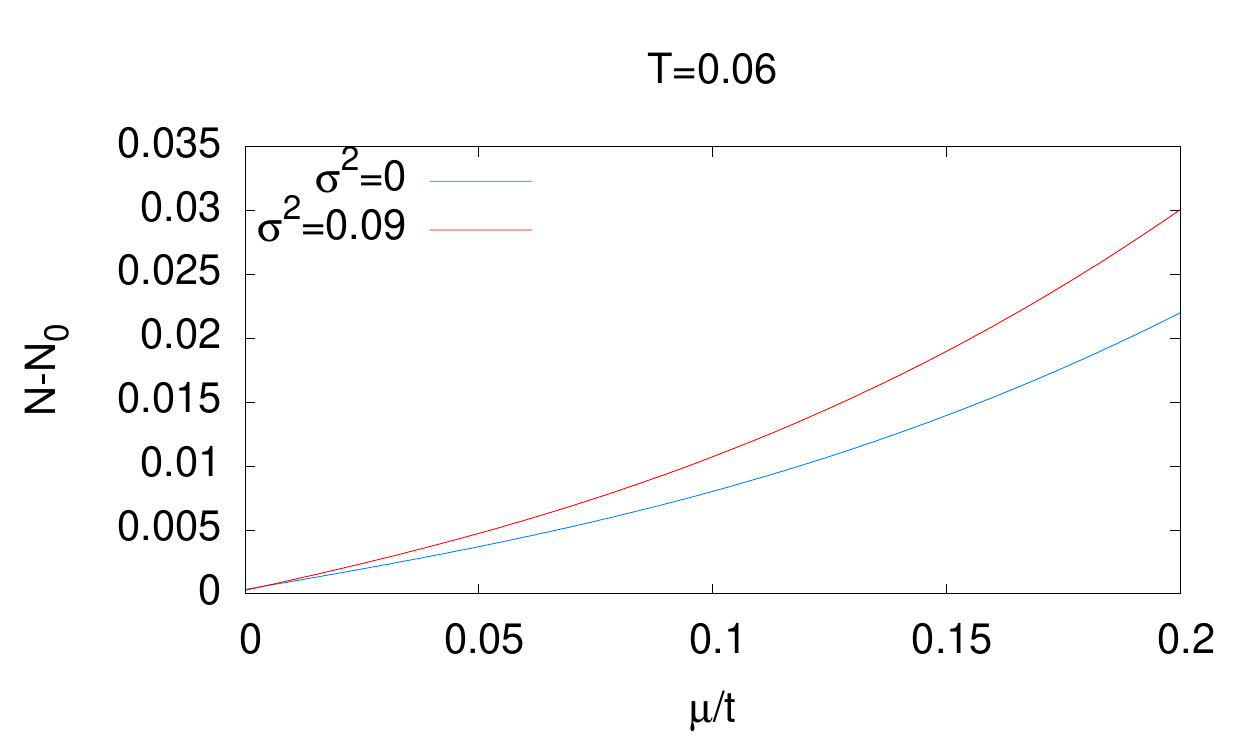}
\includegraphics[width=0.8\columnwidth]{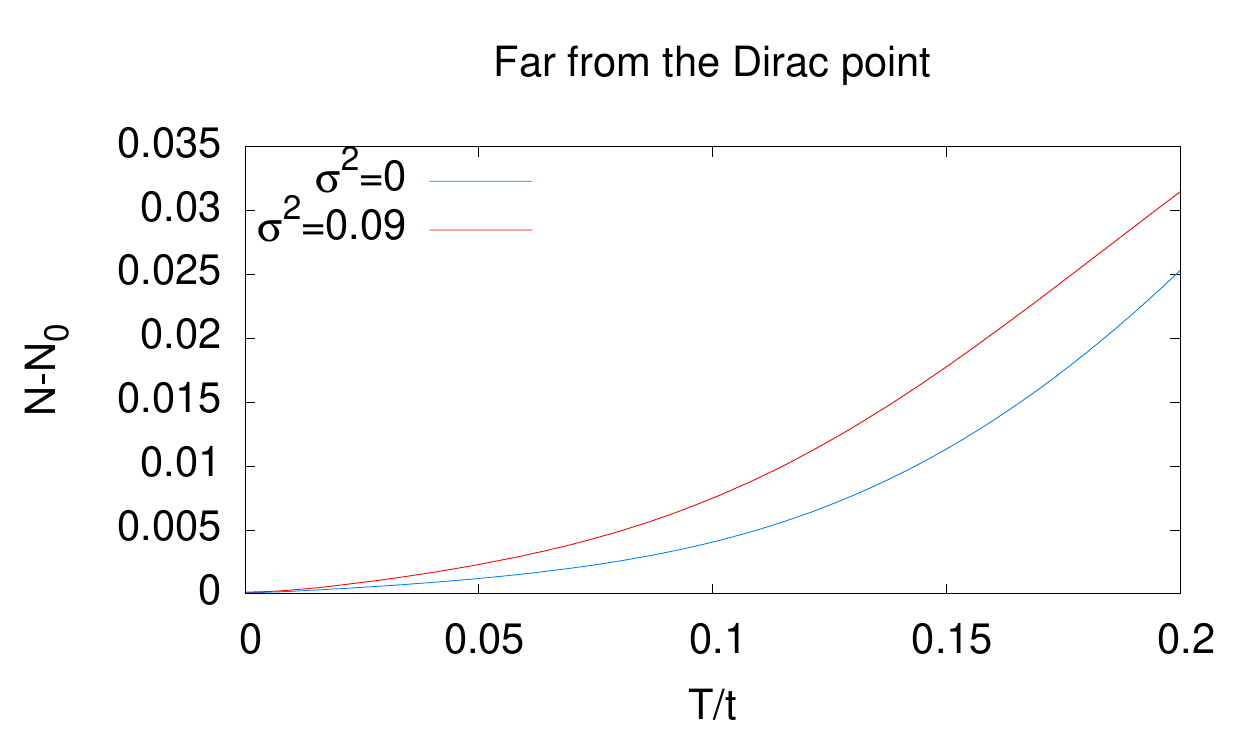}
\caption{Numerical results for the doping with Chemical potential with fixed temperature (top) and with temperature with a fixed chemical potential (bottom). The picture is in hopping units with $t^{'}=0.2t$,$n_0=2$ and $\mu_0=0.6t$}
\label{particleself1}
\end{center}
\end{figure}

In the previous section we have seen that the effects induced by the asymmetry are quite relevant from an intellectual point of view, but when we use the real parameters of graphene we notice a small effect. We have also seen that the effects  the disorder caused the shift of the DP and they increase the effects of the asymmetry. For this reason we can perform a comparison between the numerical results of the tight binding integration with and without interaction. In particular we can analyze the effect with $\mu=0$ and at high temperature with a minimum number of electrons in a band equal to $n_0=2$ and the maximum equal to $n=4$.\\
We can first analyze the  $\mu=0$  case. In the DP the disorder increases the effect of the thermal doping, as we can see from fig.(\ref{particleself0}). In particular we can obtain an analytical approximation of the  amplified thermal doping fitting the red curve in the figure, obtaining the function:
\begin{equation}
\label{distermfit}
f(T)=2+aT^{5/2}
\end{equation}
where $a=(1.532\pm0.005)$ in hopping units. We can see a comparison between eq.(\ref{muzero}) and eq.(\ref{distermfit}) also in fig.(\ref{particleself0}), with a net increase. In particular the expression eq.(\ref{distermfit}) may be helpful for a quantitative comparison with the experimental datas. Furthermore if we consider the thermal doping at room temperature ($T=300K$) we obtain the following result with and without disorder:
\begin{itemize}
\item half filling($T=300 K$): \\ 
 $n \sim 3.62\mbox{ } 10^9 \mbox{ }\frac{particles}{cm^2}$;
\item half filling($T=300 K$) with disorder $\sigma^2=0.09$:   \\
$n \sim 2.27\mbox{ } 10^{10}\mbox{ }\frac{particles}{cm^2}$
\end{itemize}
So with the disorder we can increase the thermal doping by one order of magnitude. 
%
It is interesting also to analyze the effect of the disorder at high temperature. In particular in this case we can analyze the number of particles as a function of temperature and also the chemical potential and as we can see in fig.(\ref{particleself1}) in both cases the disorder can increase the effect of the thermal doping. These properties might be important not only for the physics of graphene, but also for interesting application as a semiconductor. In particular we did not report the low temperature limit because it is less important for application.  Furthermore considering that the single particle solution is it valid only for a restricted value of temperatures, by eq.(\ref{tempdis}), we know the low temperatures correspond also to low values of the disorder, so for having significative differences between the disorder and the ordered case we should go beyond physical values of the  $\sigma^2 \simeq 0.1$.

\section{Compressibility}
The compressibility of the Fermi gas is a well known property which can be measured by means of a Scanning Tunnelling Microscopy (STM). In fact thanks to the STM we can have maps of two dimensional conductors and semiconductors and we can also measure the DOS. We can define the compressibility as:
\begin{equation}
K=\left(\frac{\partial n}{\partial \mu}\right)_{T}
\end{equation}
where there is a very high sensibility for measuring the charge fraction, as reported in \cite{Yoo25041997}. If we know the voltage between the tip and the sample we can relate it to the inverse compressibility as $\Delta V=\rho_g\frac{1}{e^2}\frac{\partial \mu}{\partial n}$, as reported in \cite{Martin-2007} for graphene. In particular in this way it is possible to analyze the effects of the disorder on the system. 
Furthermore it might be easier to measure the effect of the impurities and the e-ph interaction on the compressibility than the thermal doping and for a Fermionic system we can use the Einstein relation between the conductivity and the compressibility, i.e. $\label{einstein} \sigma=eD\left(\frac{\partial n}{\partial \mu}\right)$,
where $D$ is the diffusion coefficient linked to the disorder of the system, as reported in \cite{Martin-2007}.\\
We can use the functional form of eq.(\ref{gapdos}) for calculating the compressibility as $K=\frac{1}{n_0}\frac{\partial n}{\partial \mu}$, following the Seitz theorem, as reported in Mahan\cite{Mahan}. In particular inside the SCBA approximation we have seen that the DOS of the system is a local function independent of $k$ and in the case of the static disorder it is also indipendent of $\mu$, so deriving by $\mu$ and dividing the DOS in two parts respectively for electrons and holes, we shall obtain:
\begin{equation}
\begin{split}
&\chi=\frac{\partial n}{\partial \mu}=\int_{-\Lambda_{-}}^{0}DOS_{L}(\omega)\frac{\partial f(\omega+\mu)}{\partial \mu}d\omega+\\
&+\int_{0}^{\Lambda_{+}}DOS_{R}(\omega)\frac{\partial f(\omega+\mu)
}{\partial \mu}d\omega
\end{split}
\end{equation}
that is an exact expression for a disordered system. In particular if we consider just the static disorder, we will have to include just the first order correction and we can neglect the Hartree contribution, which shifts the energy to a density dependent value $\Sigma_{H}(n)$, dependent on $\mu$ only, as reported in eq.(\ref{selfloc}). In particular the propagator of eq.(\ref{propgreen}) will present the additional correction of $\Sigma_{H}(n)$.
The total compressibility of the system can be written as:
\begin{equation}
\chi=\int d \omega DOS(\omega) \frac{\partial f(\omega -\mu^{'})}{\partial \mu^{'}}\left( 1-\frac{\partial \Sigma_{H}}{\partial \mu}\right)
\end{equation}
where $\mu^{'}=\mu-\mu_0-\Sigma_{H}$, $\mu_0=3t^{'}-Re{\Sigma}$ is the DP shift. So we can write $\Sigma_{H}=-\alpha \chi$ and the compressibility will be:
\begin{equation}
\label{RPA}
\chi=\frac{\chi_0}{1-\alpha\chi_0}
\end{equation}
where $\chi$ is the total compressibility and $\chi_0$ is the compressibility with just the Hartree correction. we recognize in eq.(\ref{RPA}) a structure similar to the RPA. In particular in the next subsection we shall analyze the simple impurity scattering correction and then the e-ph contribution.
\subsection{Hartree contribution}
We can calculate the compressibility from the impurity scattering considering the expression of the DOS reported in fig.(\ref{confrontoselfdos2}) and fitting the expression with an easier form of the DOS, expressed as a polynomial expansion and with a non zero value in the DP, as:
\begin{equation}
DOS(\omega)=\sqrt{\Gamma^2+\alpha_{\sigma}^2 \omega^2}\pm\beta_{\sigma R/L}\omega^2
\end{equation}
in this way the compressibility will be made by a contribution from the first order and another one from the second order, considering that the derivative by $\mu$ is equal to a derivative by $\omega$ with an opposite sign and with $L/R$ distinction between electrons and holes. In particular also in this case there will be a behavior near the DP and another one far from the DP. In particular for the low temperature result we can use again the Sommerfeld expansion for the first order as:
\begin{equation}
\begin{split}
\partial_{\mu^{'}}n_1\simeq DOS(\mu^{'})+\frac{\pi^2}{6}DOS^{''}(\mu^{'})T^2 \Rightarrow\\
\partial_{\mu^{'}}n_1 \simeq \sqrt{\Gamma^2+(\alpha_{\sigma} \mu^{'})^2}+\frac{\pi^2}{6}T^2\frac{\alpha^2 \Gamma^2}{(\Gamma^2+(\alpha_{\sigma}^2 \mu^{'})^2)^{\frac{3}{2}}}
\end{split}
\end{equation}
and for the contribution of the second order we will have:
\begin{equation}
 \label{comp2part}
 \begin{split}
&\partial_{\mu^{'}}n_2=\\
&\pm \beta_{R/L}\int_{-\infty}^{\frac{\mu^{'}}{T}}dx (-\frac{\partial f(x)}{\partial x})((xT)^2+(\mu^{'})^2+2xT\mu^{'})\\
&\Rightarrow \partial_{\mu^{'}}n_2=\pm \beta_{R/L}((\mu^{'})^2+\frac{\pi^2}{3}T^2)
\end{split}
\end{equation}
and so the total  low temperature compressibility is:
\begin{equation}
\label{bassatemppos}
\begin{split}
&\partial_{\mu^{'}}n=\sqrt{\Gamma^2+(\alpha_{\sigma} \mu^{'})^2}+\frac{\pi^2}{6}T^2\frac{\alpha_{\sigma}^2 \Gamma^2}{(\Gamma^2+(\alpha_{\sigma} \mu^{'})^2)^{\frac{3}{2}}}+\\
&\pm \beta_{R/L}\left( (\mu^{'})^2+\frac{\pi^2}{3}T^2 \right)
\end{split}
\end{equation}
We notice that for $T \to 0$ the compressibility reduces to the DOS of the system.
The high temperature limit will be quite similar, but with a further correction from the integrals of the second order, because near the DP we shall consider both the contributions from electrons and holes:
\begin{equation}
\begin{split}
&\partial_{\mu^{'}}n_{2}=\\
&=\beta_{L}\int_{-\infty}^{\frac{\mu^{'}}{T}}dx (-\frac{\partial f(x)}{\partial x})((xT)^2+(\mu^{'})^2+2xT\mu^{'}) +\\
&+\beta_{R}\int_{-\frac{\mu^{'}}{T}}^{+\infty}dx (-\frac{\partial f(x)}{\partial x})((xT)^2+(\mu^{'})^2+2xT\mu^{'})
\end{split}
\end{equation}
and solving the integrals we will obtain linear corrections in temperature and cubic corrections in the chemical potential as:
\begin{equation}
\begin{split}
\partial_{\mu^{'}}n_{2}=&(\beta_R+\beta_L)(\frac{(\mu^{'})^2}{2}+T^2\frac{\pi^2}{6})+\\
&+(\beta_R-\beta_L)(2T\mu^{'}\log{(2)}+\frac{(\mu^{'})^3}{12T})
\end{split}
\end{equation}
regarding the second order, we can use a box integration in the interval $[-1,1]$ dividing the integrand by $2$. In this way the total compressibility near the DP will be:
\begin{equation}
\begin{split}
&\chi=\frac{\partial n}{\partial \mu^{'}}= \frac{\Gamma}{4T}[ (\mu+T)\sqrt{1+\frac{\alpha_{\sigma}^2}{\Gamma^2} (\mu^{'}+T)^2}-\\
&-(\mu-T)\sqrt{1+\frac{\alpha_{\sigma}^2}{\Gamma^2} (\mu^{'}-T)^2}]+\\
&+\frac{\Gamma^2}{4 \alpha T}\left[ \mbox{asinh}(\frac{\alpha_{\sigma}}{\Gamma} (\mu^{'}+T)-\mbox{asinh}(\frac{\alpha_{\sigma}}{\Gamma} (\mu^{'}-T))\right]+\\
&+(\beta_R+\beta_L)(\frac{(\mu^{'})^2}{2}+T^2\frac{\pi^2}{6})+\\
&+(\beta_R-\beta_L)(2T\mu^{'}\log{(2)}+\frac{(\mu^{'})^3}{12T})
\end{split}
\end{equation}
\subsection{Fock contribution} 
\begin{figure}[htbp]
\begin{center}
\label{compr}
\subfigure{
\includegraphics[width=0.8\columnwidth]{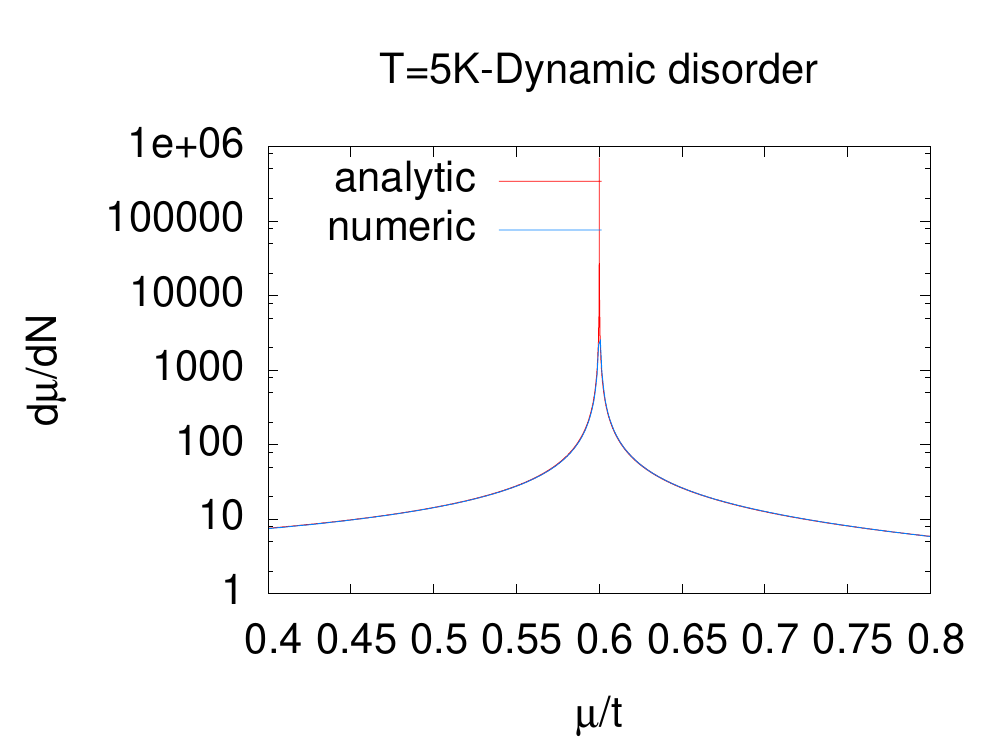}}
\subfigure{
\includegraphics[width=0.8\columnwidth]{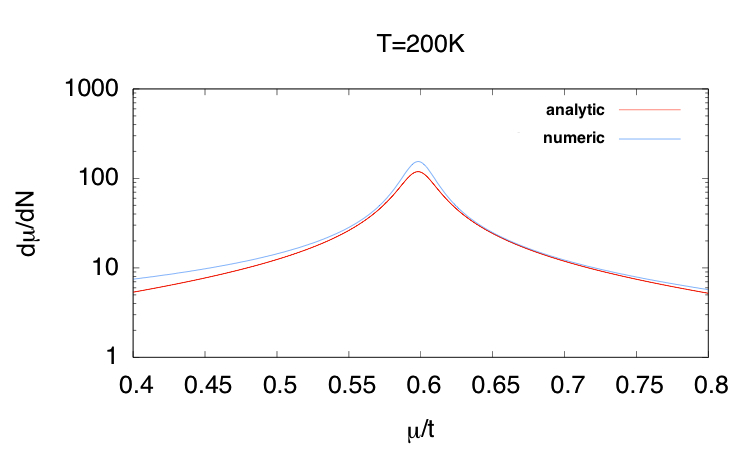}}
\caption{Quantum capacitance as a function of the chemical potential at low temperature (top) and high temperature (bottom) with a comparison between numerical and analytical calculation. In both pictures $k_b=1$, the energy scale is in hopping units  $t^{'}=0.2t$.}
\label{numbsom}
\end{center}
\end{figure}

\begin{figure}
\begin{center}
\label{lin_cond}
\includegraphics[width=0.8\columnwidth]{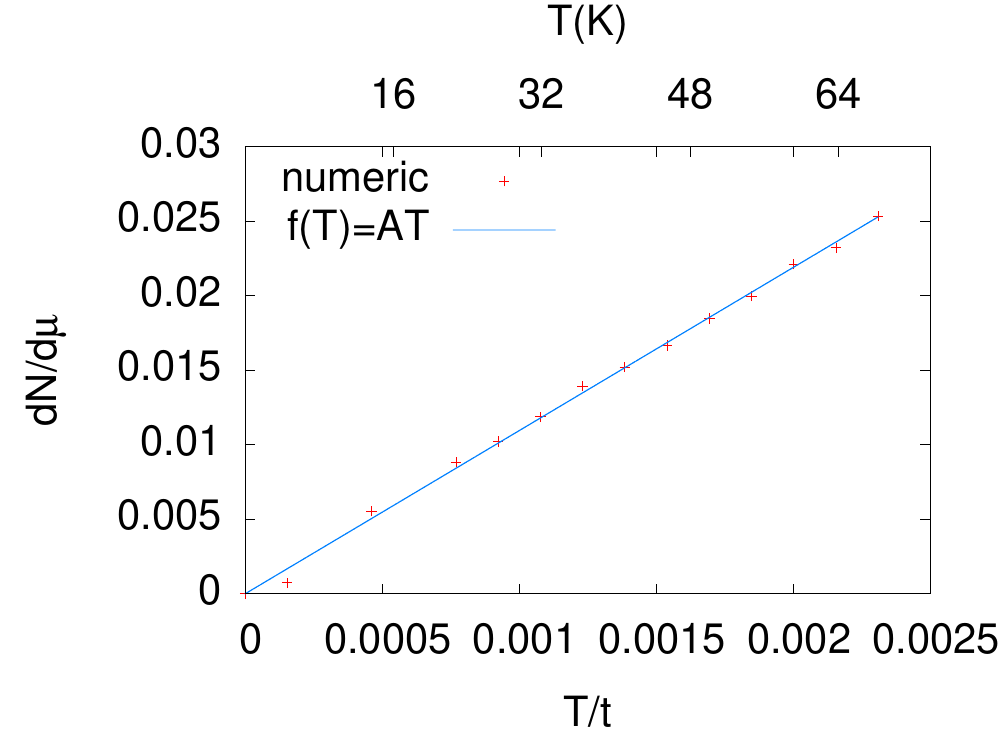}
\caption{Quantum capacitance in the DP for low temperature computed numerically for $t^{'}=0.2t$ in hopping units and in Kelvin. We notice the linear dependence, but with a deviation for a temperature under $16K$.}
\end{center}
\end{figure}

We have seen that the interactions in the system can be modeled like a RPA correction. In particular we can see that calculating $\alpha$ we will be able to increase the effect of the interaction in the system. In particular the constant $\alpha$, using the values of the deformation potential of the previous section, will be:
\begin{equation}
\begin{split}
&\alpha=2|g_{\mathbf{q}=0}|^{2}D(\mathbf{q}=0,i\nu_n=0) \\
& \alpha \simeq 1.3 (\mbox{t units})
\end{split}
\end{equation}
and so using the eq.(\ref{RPA}) we will be able to increase the effect of the asymmetry and disorder to the compressibility. We can see the results in fig.(\ref{compr}).
Furthermore considering the temperature dependence of the compressibility at the Dirac point, we can speculate on the temperature dependence of the conductivity with $\sigma=D \frac{\partial N}{\partial \mu}$, assuming a weak dependence of the diffusion coefficient from the temperature. In particular we obtain the following fitting function:
\begin{equation}
f(T)=AT
\end{equation}
where we notice a linear dependence and $A=(10.95 \pm 0.07)$. We can observe the numerical result with the fitting function in figure \ref{lin_cond}, where we notice that the points near the zero, which are the most important one, because near the values used in the experiments, they do not present an exact linear dependence, but it is not easy to verify this dependence just numerically, especially near the zero. 
\section{Conclusions}
We can finally analyze the results in the paper. In particular we have seen we took inspiration for our work from an experimental paper where they measures an anomal conductivity and furthermore introduced a new experimental set-up\cite{PhysRevLett.110.216601}. We tryied to make some basis for those unexpected phenomena. In particular we have seen that the effect of the e-h asymmetry at low energy induced an asymmetry in the DOS, which is perfectly symmetric without $t^{'}$. Moreover we have seen that the number of particles of the system are functions of the temperature just throughout $\beta$, which is the e-h correction. This kind of correction allows us to move the Fermi level of the system with temperature, but even though the property is interesting for the physics of graphene, there is a small effect on real material. This aspect has induced the study of disorder and its combination with the e-h asymmetry. The impurity scattering with the SCBA approach allows us to increase the thermal doping at room temperature by one order of magnitude and considering also the correction to the DOS, at fixed energy, we found strong effects in the DP. In particular thanks to the disorder we can shift the DP increasing the variance of the disorder and with the dynamic disorder and the e-ph interaction this shift will be temperature dependent. This property is especially evident in the calculations of the interacting DOS and the compressibility.\\
We have seen that exists an experimental measure of $t^{'}$. It may be possible to measure these properties with the Scanning Tunneling Spectroscopy (STS) and the Angle Resolved Photoemission Spectroscopy (ARPES). In particular the STS can measure the DOS of the system, for measuring the DOS and its asymmetry and with the ARPES is possible to measure the band structure of the material and in particular the e-ph coupling in the system. There are some previous studies about the effects of irradiations on graphene samples\cite{:/content/aip/journal/apl/97/17/10.1063/1.3502610}, using an electron beam with $30 Kev$, registering a shift of the DP of $\sim 20 V$ for graphene on $SiO_2$ and this is probably caused by the strong interaction with the substrate otherwise they also saw DP shift for suspended graphene, up to $\sim 0.16 V$ and it suggests that disorder may induce this property. This is a possible way for verifying these effects and also manipulating the material $ad hoc$. In fact the STM can manipulate single charges and in this way we may be able to choose and shift the insulating point according to our necessities. Furthermore as possibles perspectives we notice that we have not considered yet the  electron-electron interaction, and very recently they studied the Coulomb interaction in association to the disorder\cite{PhysRevB.91.075416}, and also the combined effect with the spin for considering short and long (RKKY) magnetic interaction and the quantum spin hall effect, because the Kane-Mele model\cite{PhysRevLett.95.226801} is  similar to the $t^{'}$ correction. Moreover the last year an experimental group from Stanford \cite{Liu16012014} discovered a three dimensional Dirac material as the $Na_3Bi$, with a band structure presenting Dirac cones. New research prospectives may be to study the e-h asymmetry for this kind of materials, originated by a next nearest neighbors in a tight-binding model. Probably thanks to the new geometries it may possible to find new thermal and magnetic effects, or strong impurities interactions, arriving to new technological applications in electronics. Furthermore really recently emerged the possibility of making honey-comb lattices of cold atoms\cite{jos}, where they obtained a DP band structure and a quantized conductivity, so it may be possible to apply similar calculations to those system. 
\section*{Acknowledgments}
This work has been done for my master thesis at the Universit\'a degli studi dell'Aquila in Italy. 
\bibliography{biblio}
\newpage
\section{Appendix 1-Dos}
Using eq.(\ref{svilrid}) the expression for calculating the DOS is:
\begin{equation}
\rho(E)=\int d\boldsymbol{q} \delta(E-E_{\pm}(q,\theta))
\end{equation}
Now using the Dirac delta function properties:
\begin{equation}
 \int_{-\infty}^{+\infty}dx\delta(f(x))=\sum_{i=1}^{N}\int_{-\infty}^{+\infty}dx\frac{\delta(x-x_i)}{|f^{'}(x_i)|}
\end{equation}
we directly obtain the $q$ integral with the delta arguments:
\begin{equation}
\label{dos3}
q_{1,2}(\epsilon,\theta)=\frac{\pm v_f \pm\sqrt{v_f^2-4A(\theta) \epsilon}}{2A(\theta)}
\end{equation}
and so:
\begin{equation}
\label{solpic}
\rho(\epsilon)=\frac{L^2}{(2\pi)^2}\int_{0}^{2\pi}d\theta \frac{q_{1}(\epsilon,\theta)}{\mp v_f+2A(\theta)q_1(\epsilon,\theta)}
\end{equation} 
Considering our interest for the DP we can expand our solutions up to the second order of $\epsilon$ and using directly eq.(\ref{svilrid}) where with the $q(\epsilon,\theta)$:
\begin{equation}
\frac{q_1(\epsilon,\theta)}{v_f}= \frac{\epsilon}{v_{f}^2}-\frac{A(\theta)}{v_{f}^2}q_1(\epsilon,\theta)^2
\end{equation}
and so thanks to eq.(\ref{dos3}) and eq.(\ref{solpic}) we can expand the integral:
\begin{equation}
\label{dosintegral}
\rho(\epsilon)=\frac{L^2}{(2\pi)^2}\int_{0}^{2\pi}d\theta \pm \left( \frac{\epsilon}{v_{f}^2}+3\frac{A(\theta)}{v_{f}^4}\epsilon^2  \right)
\end{equation}
and solving the eq.(\ref{dosintegral}), the final result will be equal to the volume of the unitary cell for the integrand function multiplied by $2\pi$, because there is just one angular dependence and it is zero. Finally the expression of the DOS without a degeneracy factor $g=4$ is;
\begin{equation}
\rho(\epsilon)=\frac{3\sqrt{3}a^2}{ 4\pi}\frac{|\epsilon|}{v_{f}^2}\pm \frac{27\sqrt{3}a^3}{8 \pi}\left(\frac{t^{'}}{t}\right)\frac{\epsilon^2}{v_{f}^3}
\end{equation}
like it has been reported in the text.
\section{Appendix 2-Doping}
\subsection{High temperature limit}
Inside this regime we can use the result of eq.(\ref{muzero}), but with $\mu \neq 0$ and $|\mu|\ll T$, obtaining some corrections to the previous behavior. Taylor Expanding the Fermi-Dirac function up to the second order:
\begin{eqnarray}
\begin{split}
&f(\epsilon,\mu,T)-f(\epsilon,0,T)+f(\epsilon,0,T)\simeq f(\epsilon,0,T)+\\
&+\left[\frac{\partial f}{\partial \mu}\right]_{\mu=0} \mu+\left[\frac{\partial^2 f}{\partial \mu^2}\right]_{\mu=0} \frac{\mu^2}{2}
\end{split}
\end{eqnarray}
and using it inside eq.(\ref{nummuzero}) we will have eq.(\ref{muzero}) plus other correction with a $\mu$ dependence:
\begin{equation}
\begin{split}
&n(\mu,T)=\\
&\int_{-\infty}^{+\infty}\rho(\epsilon)\left( f(\epsilon,0,T)+\left[\frac{\partial f}{\partial \mu}\right]_{\mu=0} \mu+\left[\frac{\partial^2 f}{\partial \mu^2}\right]_{\mu=0} \frac{\mu^2}{2} \right)d \epsilon
\end{split}
\end{equation}
and so:
\begin{equation}
\begin{split}
&n(\mu,T)-n(0,T)=\\
&\int_{-\infty}^{+\infty}\rho(\epsilon) \left( \left[\frac{\partial f}{\partial \mu}\right]_{\mu=0} \mu+\left[\frac{\partial^2 f}{\partial \mu^2}\right]_{\mu=0} \frac{\mu^2}{2} \right)d \epsilon
\end{split}
\end{equation}
 we can write the derivatives in $\mu$ as derivatives in $\epsilon$ as $\frac{\partial f(\epsilon,\mu,T)}{\partial \mu}=-\frac{\partial f(\epsilon,\mu,T)}{\partial \epsilon}$ with a consequent dependence of the functions from the energy. Considering the different contributions relative to electrons and holes and the first and second order:
\begin{equation}
\begin{split}
&n(\mu,T)-n(0,T)=\\
&-\int_{-\infty}^{0}\rho_{-}(\epsilon)\left(\left[\frac{\partial f}{\partial \epsilon}\right]_{\mu=0} \mu+\left[\frac{\partial^2 f}{\partial \epsilon^2}\right]_{\mu=0} \frac{\mu^2}{2} \right) d \epsilon-\\
&-\int_{0}^{+\infty}\rho_{+}(\epsilon)\left(\left[\frac{\partial f}{\partial \epsilon}\right]_{\mu=0} \mu+\left[\frac{\partial^2 f}{\partial \epsilon^2}\right]_{\mu=0} \frac{\mu^2}{2} \right) d \epsilon =\\
&=n_1(\mu,T)+n_2(\mu,T)
\end{split}
\end{equation} 
where $n_1(\mu,T)$ considers just contributions from first order and $n_2(\mu,T)$ from second order. Using again the substitution  $\frac{\epsilon}{T}=x$ and with a change of the sign of the integrals regarding $\rho_{-}(\epsilon)$:
\begin{equation}
n_1(\mu,T)=\mu \left(2 \alpha TI_1+2 \beta T^2I_2 \right)
\end{equation}
where every $I_n$ is given by:
\begin{equation}
\label{intrecn}
I_n=T^{n}\int_{0}^{+\infty}x^n\left(-\frac{\partial f(x)}{\partial x}\right)dx
\end{equation}
and in our case $I_1=\log{2}$ $I_2= \frac{\pi^2}{6}$.\\
In a similar way we obtain:
\begin{equation}
\begin{split}
&n_2(\mu,T)-n(0,T)=\\
&\left[\mu^2\left[ \rho_{\pm}(\epsilon)\frac{\partial f(\epsilon)}{\partial \epsilon}\right]_{0}^{+\infty}-\mu^2\int_{0}^{\infty}d\epsilon(\alpha+2\beta \epsilon)(-\frac{\partial f(\epsilon)}{\partial \epsilon})\right]
\end{split}
\end{equation}
and so neglecting exponential small contributions:
\begin{equation}
n_2(\mu,T)=-\mu^2\left(  \alpha I_0+2\beta TI_1\right)
\end{equation}
and putting together everything we obtain what has been reported below.
\section{Appendix-3}
In particular we can find an expression for the self energy considering some properties of the involved integrals:
\begin{equation}
\label{prop0}
I_n(z,\Lambda)=\int_{0}^{\Lambda} d \epsilon \frac{\epsilon^n}{z-\epsilon}
\end{equation}
where $\epsilon=\omega$($\hbar=1$) and $\omega-\Sigma(\epsilon)=z$. They present some recursive relations such as:
\begin{eqnarray}
\frac{\Lambda^n}{n}=\int_{0}^{\Lambda} d \epsilon \epsilon^{n-1}\frac{z-\epsilon}{z-\epsilon}= \\
z\int_{0}^{\Lambda} d \epsilon \frac{\epsilon^{n-1}}{z-\epsilon}-\int_{0}^{\Lambda} d \epsilon \frac{\epsilon^{n}}{z-\epsilon}
\end{eqnarray}
and so:
\begin{equation}
\label{prop1}
I_n(z,\Lambda)=zI_{n-1}(z,\Lambda)-\frac{\Lambda^n}{n}
\end{equation}
with the properties:
\begin{eqnarray}
\label{specchio}
I_n(z,\Lambda_{+})=I_n(-z,\Lambda_{-}) \\
I_n(-z,\Lambda_{+})=I_n(z,\Lambda_{-})
\end{eqnarray}
so using eq.(\ref{compactdos}) and the eq.(\ref{prop1}) we arrive to an expression for the self energy of the system:
\begin{equation}
\label{auto1}
\frac{\Sigma (\epsilon)}{n_i \sigma ^2}=\alpha G(z)+\beta  F(z)
\end{equation}
where 
\begin{eqnarray}
\label{fun_self}
G(z)= \Lambda_{-}-\Lambda_{+}+ z\ln{\frac{z^2}{(z+\Lambda_{-})(z-\Lambda_{+})}}  \\
F(z)=zG(z)-\frac{\Lambda_{+}^2+\Lambda_{-}^2}{2}
\end{eqnarray}
In particular using $z=\epsilon$ we have:
\begin{equation}
\label{noncons2}
\begin{split}
&\frac{\Sigma (\epsilon)}{n_i \sigma ^2}=(\alpha+\beta \omega) \\
&\left( \Lambda_{-}-\Lambda_{+}+ \omega\ln{\left(\frac{\omega^2}{(\omega+\Lambda_{-})(\omega-\Lambda_{+})}\right)}  \right)-\\
&-\beta \frac{\Lambda_{+}^2+\Lambda_{-}^2}{2}
\end{split}
\end{equation} 
\section{Appendix-spectral properties}
This is the full expression of the complex logarithm with the corrections:
\begin{equation}
\begin{split}
&\log{z^2}=\\
&\log{\left((\omega-\Delta(\omega))^2 -\Gamma(\omega)^2 \right)}+2i\arctan{\frac{\Gamma(\omega)}{\omega-\Delta(\omega)}}\mbox{sgn}(\omega)=\\
&\log{\rho_{1}}+2i\theta_{1}\\
&-\log{z+\Lambda_{-}}=\\
&-\frac{1}{2}\log{(\omega-\Delta(\omega)+\Lambda_{-})^2-\Gamma(\omega)^2}-i\mbox{sgn}(\omega)(\frac{3\pi}{2}+\\
&+\arctan{\frac{\Gamma(\omega)}{\omega-\Delta(\omega)+\Lambda_{-}}})=\\
&-\frac{1}{2}\log{\rho_2}-i\theta_{2}\\
&-\log{z-\Lambda_{+}}=\\
&-\frac{1}{2}\log{(\omega-\Delta(\omega)-\Lambda_{+})^2-\Gamma(\omega)^2}-i\mbox{sgn}(\omega)(-\frac{\pi}{2}+\\
&+\arctan{\frac{\Gamma(\omega)}{\omega-\Delta(\omega)-\Lambda_{+}}})=\\
&-\frac{1}{2}\log{\rho_3}-i\theta_{3}
\end{split}
\end{equation}
and $\Gamma(\omega)$ and $\Delta (\omega)$ are the the expressions reported in eq.(\ref{analiticself}). 
\end{document}